\documentclass{article}

\usepackage{microtype}
\usepackage{graphicx}
\usepackage{subcaption}
\usepackage{booktabs} 

\usepackage{xurl}
\usepackage{hyperref}

\usepackage[preprint]{icml2026}

\usepackage{amsmath}
\usepackage{amssymb}
\usepackage{mathtools}
\usepackage{amsthm}

\usepackage{multirow}
\usepackage{booktabs}
\usepackage{tabularx}
\usepackage[table]{xcolor}
\usepackage{color}

\usepackage[capitalize,noabbrev]{cleveref}

\theoremstyle{plain}

\theoremstyle{definition}

\theoremstyle{remark}

\usepackage[textsize=tiny]{todonotes}

\icmltitlerunning{GA-Field: Geometry-Aware Vehicle Aerodynamic Field Prediction}

\begin{document}

\twocolumn[
  \icmltitle{GA-Field: Geometry-Aware Vehicle Aerodynamic Field Prediction}



  \icmlsetsymbol{corr}{*}

  \begin{icmlauthorlist}
    \icmlauthor{Zhenhua Zheng}{sch,casia}
    \icmlauthor{Lu Zhang}{sch,casia,corr}
    \icmlauthor{Junhong Zou}{sch,casia}
    \icmlauthor{Shitong Liu}{casia}
    \icmlauthor{Zhen Lei}{sch,casia}
    \icmlauthor{Xiangyu Zhu}{sch,casia}
    \icmlauthor{Zhiyong Liu}{sch,casia,corr}

  \end{icmlauthorlist}

  \icmlaffiliation{sch}{School of Artificial Intelligence, University of Chinese Academy of Sciences, Beijing, China}
  \icmlaffiliation{casia}{MAIS, Institute of Automation, Chinese Academy of Science, Beijing, China}

  \icmlcorrespondingauthor{Lu Zhang}{lu.zhang@ia.ac.cn}
  \icmlcorrespondingauthor{Zhiyong Liu}{zhiyong.liu@ia.ac.cn}

  \icmlkeywords{Machine Learning, ICML}

  \vskip 0.3in
]



\printAffiliationsAndNotice{}  

\begin{abstract}

 Accurate aerodynamic field prediction is crucial for vehicle drag evaluation, but the computational cost of high-fidelity CFD hinders its use in iterative design workflows. While learning-based methods enable fast and scalable inference, accurately aerodynamic fields modeling remains challenging, as it demands capturing both long-range geometric effects and fine-scale flow structures. Existing approaches typically encode geometry only once at the input and formulate prediction as a one-shot mapping, which often leads to diluted global shape awareness and insufficient resolution of sharp local flow variations. To address these issues, we propose GA-Field, a Geometry-Aware Field prediction network that introduces two complementary design components: (i) a global geometry injection mechanism that repeatedly conditions the network on a compact 3D geometry embedding at multiple stages to preserve long-range geometric consistency, and (ii) a coarse-to-fine field refinement strategy to recover sharp local aerodynamic details. GA-Field achieves new state-of-the-art performance on ShapeNet-Car and the large-scale DrivAerNet++ benchmark for surface pressure, wall shear stress, and 3D velocity prediction tasks, while exhibiting strong out-of-distribution generalization across different vehicle categories.

\end{abstract}

\begin{figure}[t]
 
    \centering
    \includegraphics[width=1\linewidth]{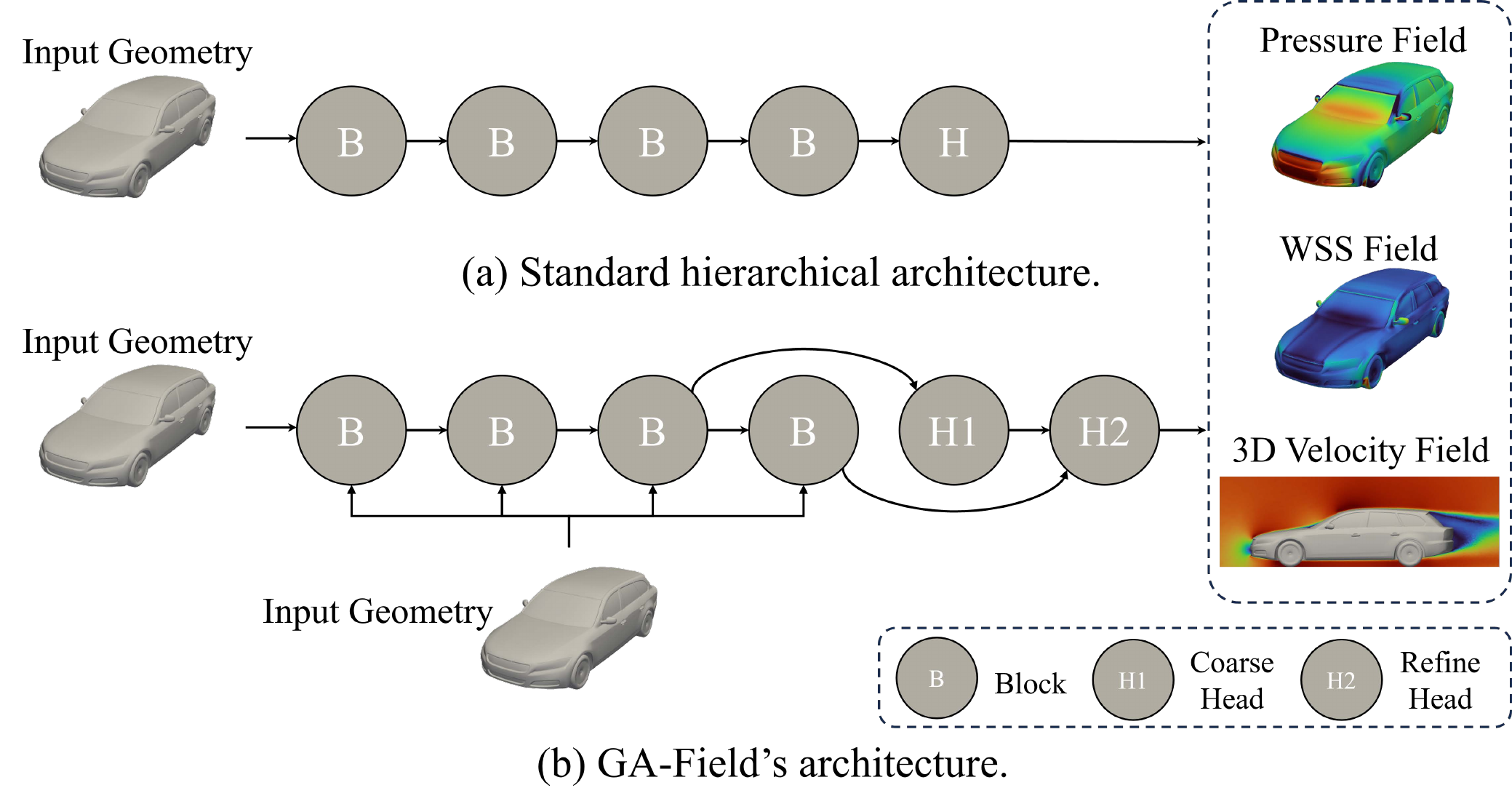}
    \caption{Architecture comparison. GA-Field differs from the standard encode once pipeline by injecting geometry across stages and refining the output instead of predicting them in one shot.}
    \label{fig:difference}
\end{figure}

\section{Introduction}

High-fidelity computational fluid dynamics (CFD) is crucial for vehicle aerodynamic design \cite{gao2025accurate}, as it provides detailed resolution of surface pressure, wall shear stress, and surrounding velocity fields that determine aerodynamic loads. However, its computational cost remains a major bottleneck for iterative optimization and rapid design exploration \cite{elrefaie2024drivaernet++}. Recently, with the rapid advances in deep learning, neural networks have become a practical alternative to CFD. Given 3D vehicle geometry under specific flow conditions (e.g., wind speed), they can predict aerodynamic fields within seconds during inference. However, achieving accurate prediction with such learning-based models remains non-trivial, as it demands capturing the underlying physical structure of aerodynamic flows.

As a key conditioning input, the 3D vehicle geometry plays a central role in determining aerodynamic behavior. This importance is also grounded in aerodynamics: the flow field is globally coupled through the governing equations and shaped by the vehicle geometry, hence even local geometric perturbations can affect the flow over much larger spatial regions \cite{WALLACE2025106177}. But the strong global dependence on geometry is often underemphasized in learning-based models. For example, most existing methods \cite{wu2024Transolver,chen2025tripnet} encode the vehicle geometry only \textit{once} at the input stage, then propagate it implicitly through the deep, multi-stage architectures. As a result, global geometric information can be progressively diluted, which weakens the model’s ability to capture long-range geometric effects.

On the other hand, sharp variation structures of the vehicle geometry (e.g., side mirrors) also strongly shape how pressure and shear are distributed over the vehicle surface, especially in near-wall boundary layers and separated wakes \cite{URQUHART2018308}. To accurately resolve these fine-scale effects, traditional CFD solvers rely on iterative error reduction such as multigrid \cite{WESSELING2001311, YAN2007445} and defect correction \cite{jayasankar2022defect}. However, existing learning-based methods typically formulate field prediction as a \textit{one-shot} mapping from geometry and flow conditions to target fields \cite{elrefaie2024surrogate, zou2026adafield}. As a result, they often struggle to reduce errors consistently across spatial scales, which limits the ability to accurately resolve fine-scale flow structures \cite{chen2025tripnet}.

To address these challenges, we propose GA-Field, a Geometry-Aware neural network for vehicle aerodynamic field prediction that jointly models long-range geometric effects and sharp local flow details. Figure~\ref{fig:difference} shows the differences between GA-Field and the standard hierarchical architecture. First, we argue that the embedding of 3D vehicle geometry should be injected into the network at multiple stages rather than being encoded only once at the input. This is consistent with recent rapid advances in generative models \cite{Du_2025_CVPR} and 3D vision \cite{Huang_2021_ICCV,jun2023shapegeneratingconditional3d}, where repeated conditioning has demonstrated strong effectiveness and emerged as an important design paradigm \cite{rebain2023attention}. Specifically, we propose a global geometry injection module to maintain persistent global shape awareness in hierarchical networks. It encodes the full vehicle geometry into a compact embedding and injects it into each network stage via FiLM modulation \cite{perez2018film}, thus preventing the conditioning signal from being diluted and improving long-range consistency across scales. Second, to better capture regions with sharp spatial variation, we introduce a field detail refinement module that follows a coarse-to-fine strategy. A coarse head predicts a low-resolution field to capture the global pattern, which is then upsampled and refined by a dedicated refinement head via residual correction. Overall, the proposed GA-Field achieves state-of-the-art performance on ShapeNet-Car~\cite{shapnetcar} and the large-scale DrivAerNet++~\cite{elrefaie2024drivaernet++} benchmark. It consistently outperforms previous methods on multiple aerodynamic tasks, including surface pressure, wall shear stress, and 3D velocity prediction, while demonstrating strong out-of-distribution (OOD) generalization across different vehicle categories. Our main contributions are summarized as
follows:

\begin{itemize}

    \item We highlight the importance of repeated geometric conditioning in learning-based aerodynamic field prediction and propose a global geometry injection module to preserve long-range geometric effects.
   
    \item We propose a coarse-to-fine field detail refinement strategy that overcomes the limitations of one-shot learning-based prediction and enables accurate resolution of fine-scale aerodynamic structures.
 
    \item GA-Field achieves a new state-of-the-art performance on ShapeNet-Car and large-scale DrivAerNet++ benchmarks across various tasks, including surface pressure, wall shear stress, and 3D velocity, together with strong OOD generalization across different vehicle categories. 
    
\end{itemize}

\begin{figure*}[!t]
 
    \centering
    \includegraphics[width=1\linewidth]{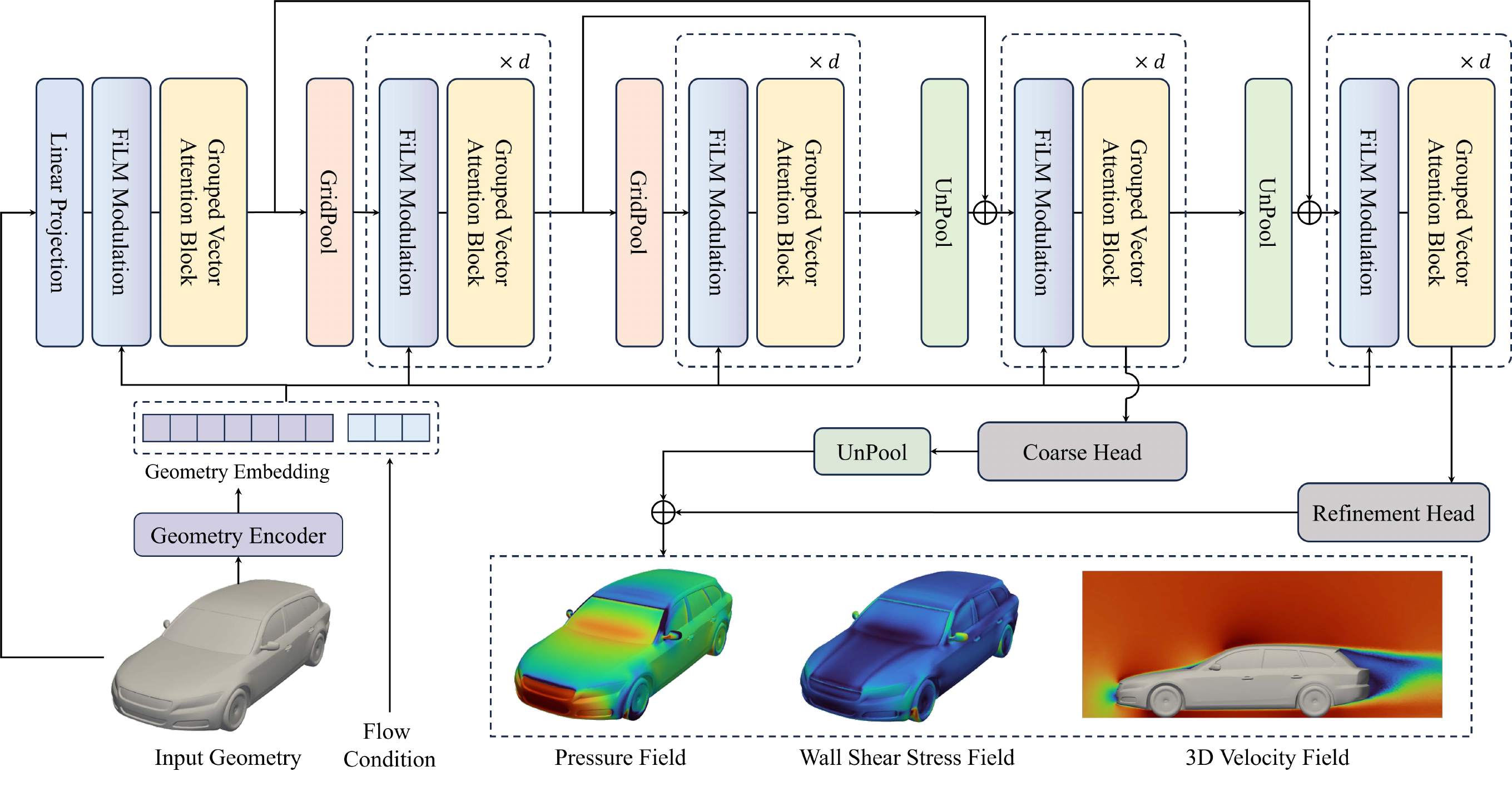}
    \caption{Overall architecture of GA-Field. GA-Field takes a vehicle geometry point cloud and flow conditions as input. A geometry encoder extracts a global embedding, which is injected into each attention block. The network outputs a coarse prediction at lower resolution and refines it at full resolution to obtain the final field.}
    \label{fig:method}
\end{figure*}

\section{Related Work}

\subsection{Aerodynamic Field Prediction}

Aerodynamic field prediction aims to map geometry and flow conditions to spatially distributed aerodynamic quantities, such as surface pressure and velocity fields. Previous works primarily differ in how the vehicle geometry is represented, which can be broadly categorized into discrete (e.g., graph and point cloud) and implicit representations.

\textbf{Discrete representation.} MeshGraphNet \cite{pfaff2021learning} learns local interactions via multi-step message passing on mesh graphs, thereby predicting next-step state updates and enabling fast approximation. X-MeshGraphNet \cite{nabian2024x} is a scalable extension of MeshGraphNet, which builds a multi-resolution graph representation to enlarge the receptive field. ReDGCNN \cite{elrefaie2024drivaernet} constructs dynamic kNN graphs based on sampled points and applies edge convolutions to aggregate local geometric relations for prediction. Alongside graphs, representative point-based models, such as PointNet \cite{qi2017pointnet} and PointMAE \cite{pang2023masked}, have recently received considerable attention in aerodynamic analysis \cite{elrefaie2025carbench}. Transolver \cite{wu2024Transolver} proposes Physics Attention, which partitions mesh points into learnable slices and computes attention only among these slices to model physical correlations. Building upon Transolver, Transolver++ \cite{luo2025transolver++} further enhances accuracy and scalability for complex PDEs on large-scale meshes. Very recently, AdaField \cite{zou2026adafield} pretrains the model on large-scale datasets and proposes a Flow-Conditioned Adapter to improve generalization in sub-domains with limited data.

\textbf{Implicit representation.} FigConvNet \cite{choy2025factorized} employs factorized implicit grids to approximate high-resolution 3D domains and uses a two-dimensional reparameterization to enable more efficient global convolutions. TripNet \cite{chen2025tripnet} encodes 3D space with three mutually orthogonal feature planes. For any query point, it projects the point onto these planes and decodes target quantities using a lightweight MLP.

Despite differences in geometry representations, many methods still adopt an encode-once, one-shot paradigm, where global geometric information is hard to preserve across scales, and fine-scale corrections are not explicitly modeled.

\subsection{Geometry-Aware Designs}

Geometry plays a central role in aerodynamics, and prior work has explored various mechanisms to incorporate geometric information into neural PDE solvers. For example, to learn PDE operators that generalize across varying domains, GINO \cite{NEURIPS2023_70518ea4} makes geometry an explicit input by encoding each shape with surface points and a signed distance function (SDF), and uses this representation to lift irregular meshes into a uniform latent grid for global FNO \cite{li2021fourier} modeling and to decode back to arbitrary query points. Similarly, to improve 3D PDE solving on arbitrary geometries, 3D-GeoCA \cite{ijcai2024p640} extracts a latent geometry representation from boundary point clouds and uses it to condition the network, instead of forcing the network to learn a difficult geometry to uniform grid mapping. SpiderSolver \cite{qi2025spidersolver} introduces spiderweb tokenization, where boundary spectral clustering and SDF distance bins partition an irregular domain into geometry-aligned tokens, with optimal-transport alignment to keep token semantics consistent across varying shapes, thereby efficiently solving PDEs on complex geometries. 

Although these methods incorporate geometry-aware designs, they are primarily motivated by adapting to varying geometric shapes and do not place additional emphasis on fine-scale detail regions. In contrast, we highlight the importance of the overall shape of each given geometry while explicitly attending to local fine detail regions.

\section{Methodology}

The overall architecture of the proposed GA-Field, is illustrated in Figure \ref{fig:method}. The inputs of GA-Field are a vehicle geometry point cloud $\mathcal{P}=\{(\mathbf{x}_i,\mathbf{f}_i)\}_{i=1}^N$ and a flow condition vector $\mathbf{c} \in \mathbb{R}^c$, where $\mathbf{x}_i\in \mathbb{R}^3$ is the 3D coordinate, $\mathbf{f}_i$ represents point-wise features (e.g., surface normals) and $c$ is the number of flow condition parameters. GA-Field aims to predict a target quantity $\hat{\mathbf{y}}_{i}$ for each point. Concretely, for vehicle surface points, we predict pressure $p_i\in\mathbb{R}$ and wall shear stress (WSS) $\mathbf{\tau}_i\in \mathbb{R}^3$. For exterior flow domain points around the vehicle, we predict the velocity $\mathbf{u}_i \in \mathbb{R}^3$.

\subsection{Backbone Architecture}
\label{backbone}

GA-Field adopts a U-Net style \cite{ronneberger2015u} point-based backbone to enable multi-scale feature learning. In previous designs \cite{zhao2021point,zou2026adafield}, the multi-scale point features are typically constructed using farthest point sampling (FPS) \cite{eldar1997farthest}, which can be computationally expensive and sensitive to non-uniform sampling \cite{wu2022point}. We therefore adopt grid pooling \cite{wu2022point} for efficient and spatially regular downsampling. Specifically, let $\mathcal{P}^0$ and $s_0$ denote the point cloud and corresponding grid size, grid pooling is defined as
\begin{equation}
\left(\mathcal{P}^{1}, \pi^{0}\right) = \operatorname{GridPool}\left(\mathcal{P}^{0} ; s_{0}\right),
\end{equation}
where $\mathcal{P}^1=\{(\mathbf{x}^1_j,\mathbf{f}^1_j)\}_{j=1}^M$ denote the downsampled point cloud and $\pi^{0}:\{1,\cdots,N\}\rightarrow\{1,\cdots,M\}$ maps each original point to its grid cluster. For feature aggregation, we leverage the grouped vector attention \cite{wu2022point} for improved scalability. Compared to standard vector attention \cite{zhao2021point} whose weight encoding layers grow with channel dimension $C$, this design partitions the value vector $\mathbf{v} \in \mathbb{R}^C$ into $g$ groups $(1\leq g\leq C)$ and predicts only $g$ shared attention weights:
\begin{equation}
\begin{gathered}
\alpha_{ij}=\omega(\gamma(\mathbf{q}_i,\mathbf{k}_j)) \\
\mathbf{f}_i^{attn}=\sum_{j\in\mathcal{N}(i)}\sum_{l=1}^{g}\sum_{m=1}^{C/g}\operatorname{Softmax}(\alpha_i)_{jl}\mathbf{v}_j^{lC/g+m},
\end{gathered}
\end{equation}
where $\gamma$ is a relation function, $\omega:\mathbb{R}^C\rightarrow \mathbb{R}^{g}$ is the learnable grouped weight enconding, and $\mathcal{N}(i)$ denotes grid cluster. Together, grid pooling and grouped vector attention form an efficient backbone for large-scale point cloud field prediction, which we use throughout GA-Field.

\subsection{Global Geometry Injection}
\label{global geometry}

In aerodynamics, geometry influences the flow beyond its local neighborhood \cite{hosseini2011pressure, vreman2014projection}. Accurate point-wise prediction, therefore, requires global shape awareness. However, many hierarchical models \cite{wu2024Transolver,zou2026adafield} encode geometry only once and rely heavily on local aggregation (e.g., kNN), making global shape cues difficult to preserve across scales.

To address this, we introduce a global geometry injection module that encodes the full vehicle shape into a geometry embedding and injects it throughout the network, ensuring consistent geometry awareness. Specifically, we employ a geometry encoder to extract a global geometry embedding. The encoder first applies grid pooling with grid size $s_g$ to the input point cloud, producing a set of geometry tokens that summarize the vehicle structure:
\begin{equation}
    \left(\mathcal{G}, \pi^{g}\right) = \operatorname{GridPool}\left(\mathcal{P}^{0} ; s_{g}\right),
\quad \mathcal{G}=\{(\mathbf{p}_k,\mathbf{z}_k)\}_{k=1}^K,
\end{equation}
where $\mathcal{G}$ denotes the geometry tokens. The encoder then applies self-attention on these tokens to model long-range structural interactions:
\begin{equation}
\begin{gathered}
    \mathbf{q}_{k}, \mathbf{k}_{j}, \mathbf{v}_{j} = Q\left(\mathbf{z}_{k}\right),\, K\left(\mathbf{z}_{j}\right),\, V\left(\mathbf{z}_{j}\right), \\
    \alpha_{kj} = \operatorname{Softmax}\!\Big(
\gamma_w\!\left(\mathbf{q}_{k}-\mathbf{k}_{j}+\gamma_p(\mathbf{p}_k-\mathbf{p}_j)\right)
\Big)_j,\\
\tilde{\mathbf{z}}_{k} = \sum_{j \in \mathcal{N}(k)} \alpha_{kj}\odot\left(\mathbf{v}_{j}+\gamma_p(\mathbf{p}_k-\mathbf{p}_j)\right),
\end{gathered}
\end{equation}
where $Q$, $K$, and $V$ are linear projection functions, $\gamma_w$ and $\gamma_p$ are two-layer MLPs. These tokens are finally aggregated by mean pooling to form a single global geometry embedding:
\begin{equation}
    \mathbf{g} = \frac{1}{\left|\mathcal{G}\right|} \sum_{k \in \mathcal{G}} \tilde{\mathbf{z}}_{k}.
\end{equation}
We treat $\mathbf{g}$ as an additional flow condition and concatenate it with the original flow condition $\mathbf{c}$ to form an extended condition $\mathbf{c}^+$. This extended condition is injected into every attention block via FiLM modulation \cite{perez2018film}. Specifically, for each attention block $\ell$, an MLP predicts feature-wise shift and scale parameters from $\mathbf{c}^+$:
\begin{equation}
    \mathbf{c}^{+}=\left[\mathbf{g} ; \mathbf{c}\right],  \quad \left(\mathbf{s}^\ell,\mathbf{r}^\ell\right)= \operatorname{MLP}(\mathbf{c}^+).
\end{equation}
These parameters are then used to modulate the intermediate features of the $\ell$-th block:
\begin{equation}
\begin{gathered}
    \Delta\mathbf{f}_i^\ell = W_{\rm{out}} \left(  \left( 1+\mathbf{r}^\ell \right) \odot W_{\rm{in}}\left(\mathbf{f}_i^\ell\right) + \mathbf{s}^\ell \right), \\
\mathbf{f}_i^\ell \leftarrow \mathbf{f}_i^\ell+\Delta\mathbf{f}_i^\ell,
\end{gathered}
\end{equation}
where $W_{\rm{in}}$ and $W_{\rm{out}}$ are linear projections. By injecting $\mathbf{g}$ into each attention block, global geometry is incorporated at all resolutions, ensuring persistent geometry conditioning throughout the network.

\begin{figure*}[!t]
 
    \centering
    \includegraphics[width=1\linewidth]{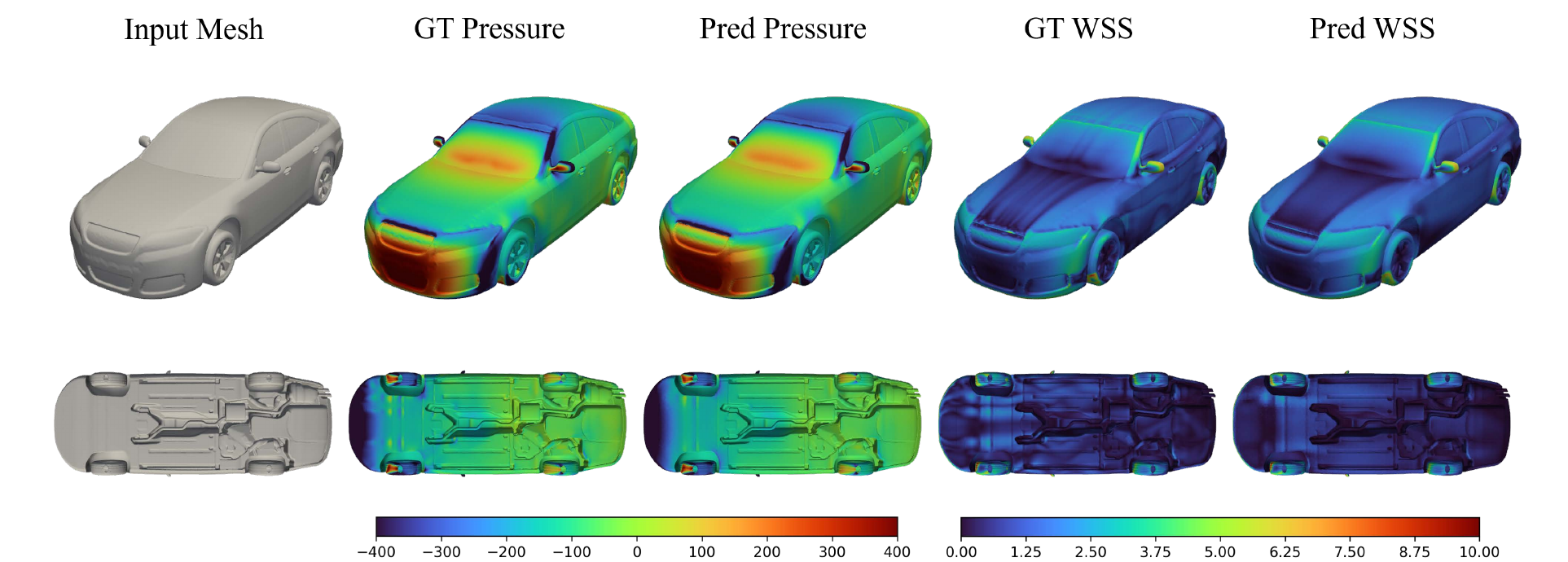}
    \caption{Comparison of pressure and wall shear stress predictions for car design \text{F\_D\_WM\_WW\_2839} (Fastback with Detailed underbody, wheels open detailed, and with mirrors) from the unseen test set. The first row shows the input surface mesh of the vehicle, along with the ground-truth pressure and wall shear stress, and the corresponding predicted pressure and wall shear stress. The second row shows the input underbody mesh, together with the ground-truth pressure and wall shear stress, and their predicted counterparts.}
    \label{fig:surface_exp}
\end{figure*}

\subsection{Field Detail Refinement}
\label{field refinement}

Aerodynamic field prediction regresses continuous physical quantities that depend strongly on accurate spatial gradients \cite{URQUHART2018308}. Therefore, errors tend to accumulate around small geometric details (e.g., side mirrors) \cite{chen2025tripnet}. Many existing models follow a one-shot prediction paradigm \cite{wu2024Transolver,chen2025tripnet}, which limits their ability to resolve fine-scale patterns.

To address this, we introduce a field detail refinement module that first captures the overall flow pattern at a coarse level, and then refines fine-scale details with a residual correction. Specifically, a coarse head operates on a lower-resolution feature representation and predicts a coarse flow field:
\begin{equation}
    \hat{\mathbf{y}}_{j}^{1} = H_{\text{coarse}}\!\left(\mathbf{f}_{j}^{1}\right),
\quad 
\hat{\mathbf{y}}_{i}^{0,\uparrow} = \hat{\mathbf{y}}_{\pi^{0}(i)}^{1},
\end{equation}
where $\hat{\mathbf{y}}_{j}^{1}$ is the coarse prediction and is upsampled via the index mapping $\pi^{0}$, yielding $\hat{\mathbf{y}}_{i}^{0,\uparrow}$. This coarse estimate captures the global structure of the flow but lacks fine-scale accuracy. A refinement head then operates at full resolution to predict a residual field that corrects fine-scale errors:
\begin{equation}
\Delta \mathbf{y}_{i}^{0} = H_{\text{refine}}\!\left(\mathbf{f}_{i}^{0}\right),
\quad \hat{\mathbf{y}}_{i}^{0} = \hat{\mathbf{y}}_{i}^{0,\uparrow} + \Delta \mathbf{y}_{i}^{0},
\end{equation}
where $\Delta \mathbf{y}_{i}^{0}$ is the residual field which corrects fine-scale variations of $\hat{\mathbf{y}}_{i}^{0,\uparrow}$ and produce the final field $\hat{\mathbf{y}}_{i}^{0}$. By decomposing prediction into a coarse estimate and a residual refinement, the model explicitly allocates capacity to correcting errors in sharp-variation regions. During training, we supervise both the final field and the upsampled coarse field using L1 loss, and use a scalar $\lambda$ to balance two terms.
\begin{equation}
\mathcal{L}=\frac{1}{N}\sum_{i=1}^N \left( \Big\| \hat{\mathbf{y}}_{i}^{0}-\mathbf{y}_i \Big\|_1 + \lambda \Big\| \hat{\mathbf{y}}_{i}^{0,\uparrow}-\mathbf{y}_i \Big\|_1 \right),
\end{equation}

\section{Experiments}

In this section, we conduct extensive experiments to evaluate GA-Field. We first describe the implementation details and experimental setup (Secs.~\ref{implementation} and \ref{exp setup}), and then report the main results on DrivAerNet++ for surface-field and 3D field prediction (Sec.~\ref{main exp}). We further assess out-of-distribution generalization across different vehicle categories on DrivAerNet++ (Sec.~\ref{ood}) and conduct ablation studies (Sec.~\ref{ablation}), including component-wise ablations and sensitivity to input point numbers. Finally, we report additional results on ShapeNet-Car (Sec.~\ref{shapenetcar}) and demonstrate practical utility via part-wise drag analysis (Sec.~\ref{part drag}).

\subsection{Implementation Details}
\label{implementation}
Following previous works \cite{chen2025tripnet,zou2026adafield}, we sample $N=32,768$ points per vehicle for training and testing. We apply grid pooling with increasing grid sizes (0.06, 0.12, 0.24, 0.48, 0.96) across scales. The geometry encoder produces $\mathbf{g}\in\mathbb{R}^{64}$ with $s_g=0.3$. We set $\lambda=0.3$ during training. The model is trained for 200 epochs with AdamW \cite{loshchilov2018decoupled} optimizer, learning rate $10^{-4}$, and cosine schedule with 3,000 warm-up iterations. We use batch size 4, weight decay 0.01. All experiments are conducted on 2 NVIDIA H100 GPUs.

\subsection{Experimental Settings}
\label{exp setup}
\noindent\textbf{Benchmark.} We evaluate our method on DrivAerNet++ \cite{elrefaie2024drivaernet++}, a large-scale, high-fidelity automotive CFD benchmark, which contains over 8,000 samples, and each vehicle surface mesh includes more than 500,000 surface points. It also provides semantic part annotations to enable part-wise analysis. For all experiments, we follow the official train/validation/test split.

\noindent\textbf{Evaluation Metrics.} We adopt the evaluation protocol used in the DrivAerNet++, including absolute errors such as mean squared error (MSE), mean absolute error (MAE), and maximum absolute error (MaxAE), as well as relative errors, including relative $\ell_2$ (Rel L2) and relative $\ell_1$ (Rel L1) errors. For pressure, we report metrics on normalized values obtained by subtracting the mean (-94.5) and dividing by the standard deviation (117.25). For WSS, we report metrics on its magnitude, computed as the $\ell_2$ norm of its three vector components. For velocity, we report both component-wise magnitude-based metrics.

\begin{table*}[!t]
  \caption{Performance comparison on surface pressure field prediction and wall shear stress field prediction across various models, including ReDGCNN \cite{elrefaie2024drivaernet}, Transolver \cite{wu2024Transolver}, FigConvNet \cite{choy2025factorized}, TripNet \cite{chen2025tripnet}, and AdaField \cite{zou2026adafield}. AdaField is specifically designed for surface pressure prediction and cannot be directly extended to wall shear stress prediction. The evaluation is conducted on the unseen test set of the DrivAerNet++ dataset.}
  \label{tab: surface results}
  \setlength{\tabcolsep}{3.1pt}
  \begin{center}
      \begin{sc}
      \begin{small}
        \begin{tabular}{l|ccccc|ccccc}
            \toprule
            \multirow{3}{*}{Method} & \multicolumn{5}{c|}{Pressure} & \multicolumn{5}{c}{Wall Shear Stress} \\
            \cmidrule(lr){2-6} \cmidrule(lr){7-11}
            
            & {MSE$\downarrow$} & {MAE$\downarrow$} & \multirow{2}{*}{MaxAE$\downarrow$} & {Rel L2} & {Rel L1} & {MSE$\downarrow$} & {MAE$\downarrow$} & \multirow{2}{*}{MaxAE$\downarrow$} & {Rel L2} & {Rel L1}\\
             
            & {($\times10^{-2}$)} & {($\times10^{-1}$)} &   & {(\%)$\downarrow$} & {(\%)$\downarrow$} & {($\times10^{-2}$)} & {($\times10^{-1}$)} &   & {(\%)$\downarrow$} & {(\%)$\downarrow$}\\
            
            \midrule
            {RegDGCNN \textsubscript{\tiny IDETC-CIE\textquotesingle24}}        & 8.29 & 1.61 & 10.81 & 27.72 & 26.21 & 13.82 & 3.64 & 11.54 & 36.42 & 36.57\\
            {Transolver \textsubscript{\tiny ICML\textquotesingle24}}      & 4.99 & 1.22 &  6.55 & 20.86 & 21.12 &  9.86 & 2.22 &  \underline{7.84} & 22.32 & 17.65\\
            {FigConvNet \textsubscript{\tiny arXiv\textquotesingle25}}      & 7.15 & 1.41 &  7.12 & 23.87 & 22.57 &  \underline{8.95} & \underline{2.06} &  8.45 & 22.49 & 17.65\\
            {TripNet \textsubscript{\tiny arXiv\textquotesingle25}}         & 5.14 & 1.25 &  6.35 & 20.05 & 20.93 &  9.52 & 2.15 &  8.14 & \underline{22.07} & \underline{17.18}\\
            {AdaField \textsubscript{\tiny AAAI\textquotesingle26}}        & \underline{4.58} & \underline{1.05} &  \underline{6.22} & \underline{19.81} & \underline{17.28} &     - &    - &     - &     - &     -\\
            \midrule
            GA-Field              & \textbf{4.13} & \textbf{1.04} &  \textbf{6.13} & \textbf{19.28} & \textbf{17.00} &  \textbf{8.18} & \textbf{1.62} &  \textbf{5.64} & \textbf{20.54} & \textbf{16.12}\\
            \bottomrule
        \end{tabular}
      \end{small}
      \end{sc}
  \end{center}
  \vskip -0.1in
\end{table*}

\begin{table}[!t]
  \caption{Performance comparison of 3D velocity field prediction on the unseen test set between TripNet \cite{chen2025tripnet} and GA-Field. Relative errors are reported in percentage (\%).}
  \label{tab: velocity results}
  \setlength{\tabcolsep}{5.5pt}
  \begin{center}
      \begin{sc}
      \begin{small}
        \begin{tabular}{l|l|cccc}
            \toprule
            \multirow{2}{*}{Metrics} & \multirow{2}{*}{Method}  & \multicolumn{4}{c}{Field} \\
            \cmidrule(lr){3-6}
             & &  {$U_x$} & {$U_y$} & {$U_z$} & {$U$} \\
            \midrule

            \multirow{2}{*}{MSE} & TripNet & \underline{6.69} &  \underline{2.24} &  \underline{2.39} &  \underline{6.71} \\
                                 & GA-Field   & \textbf{6.09} &  \textbf{1.98} &  \textbf{1.86} &  \textbf{6.08} \\
            \midrule
            \multirow{2}{*}{MAE} & TripNet &  \underline{1.49} &  \underline{0.86} &  \underline{0.86} &  \underline{1.52} \\
                                 & GA-Field   &  \textbf{1.30} &  \textbf{0.69} &  \textbf{0.67} &  \textbf{1.31} \\
            \midrule
            \multirow{2}{*}{MAxAE} & TripNet & \underline{31.31} & \underline{25.06} & \underline{22.07} & \textbf{28.09} \\
                                   & GA-Field   & \textbf{29.78} & \textbf{22.96} & \textbf{21.76} & \underline{28.27} \\
            \midrule
            \multirow{2}{*}{Rel L2} & TripNet & \underline{10.71} & \textbf{35.34} & \underline{36.39} & \underline{10.39} \\
                                    & GA-Field   &  \textbf{9.91} & \underline{35.94} & \textbf{33.79} &  \textbf{9.65} \\
            \midrule
            \multirow{2}{*}{Rel L1} & TripNet &  \underline{7.15} &  \underline{28.97} & \underline{31.12} & \underline{6.88} \\
                                    & GA-Field   &  \textbf{5.95} & \textbf{27.09} & \textbf{26.95} &  \textbf{5.76} \\

            \bottomrule
        \end{tabular}
        \end{small}
      \end{sc}
  \end{center}
  \vskip -0.1in
\end{table}

\noindent\textbf{Baselines.} We compare GA-Field with recent state-of-the-art models, covering major modeling paradigms: the implicit convolutional model FigConvNet \cite{choy2025factorized}; the GNN-based RegDGCNN \cite{elrefaie2024drivaernet}; the transformer-based models, Transolver \cite{wu2024Transolver} and AdaField \cite{zou2026adafield}; and the triplane-based model, TripNet \cite{chen2025tripnet}.

\subsection{Main Results}
\label{main exp}
\noindent\textbf{Setup.} For surface-field prediction, we use point coordinates $\mathbf{x}_i$ and surface normals $\mathbf{n}(\mathbf{x}_i)$ as input features, and add an explicit \textit{inflow-angle} cue by computing $\cos\theta=\langle \mathbf{n}(\mathbf{x}_i),\mathbf{d}_{\infty}\rangle$, where $\mathbf{d}_{\infty}$ denote the inflow direction, yielding $\mathbf{f}_i\in\mathbb{R}^7$. For 3D flow-field prediction, surface normals are unavailable for flow-domain points. We instead use an \textit{initial-velocity} feature: \textbf{0} for vehicle surface points and inflow velocity $\mathbf{u}_{\infty}$ for flow-domain points, yielding $\mathbf{f}_i\in\mathbb{R}^6$. These features provide physics priors on inflow direction and boundary conditions, improving discrimination between the vehicle and surrounding flow and enhancing accuracy.

\begin{figure*}[!t]
 
    \centering
    \includegraphics[width=1\linewidth]{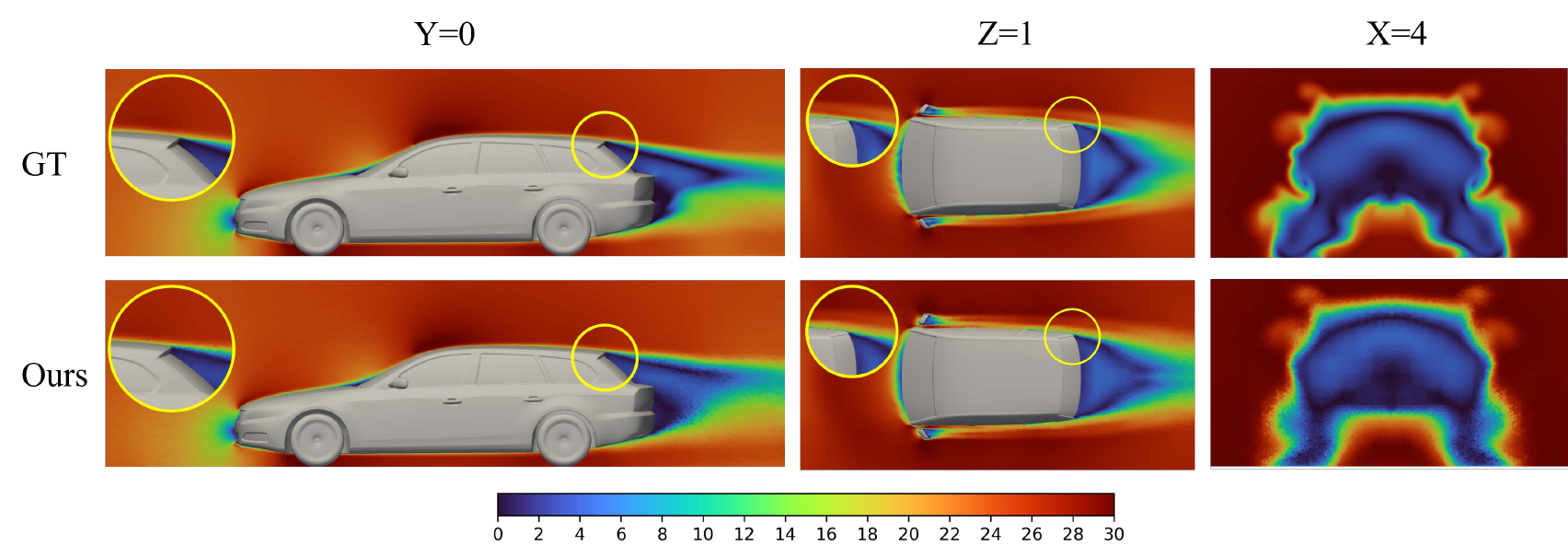}
    \caption{Comparison of 3D velocity magnitude field for the car design E\_S\_WWC\_WM\_094 (estateback with smooth underbody, with wheels closed, and with mirrors) from the unseen test set. The visualization includes three planes: y = 0 (symmetry plane), z = 1, and x = 4 (wake of the car). The first row shows the ground truth CFD results, and the second row displays our model's predictions.}
    \label{fig:volume exp}
\end{figure*}

\noindent\textbf{Results.} As summarized in Table \ref{tab: surface results} and Table \ref{tab: velocity results}, our method achieves state-of-the-art performance across surface pressure, wall shear stress (WSS), and 3D velocity field prediction. For surface pressure, GA-Field surpasses the previous pressure-specific SOTA, AdaField, reducing MSE by 9.8\% and consistently improving MAE, MaxAE, and relative errors. On WSS, GA-Field outperforms FigConvNet with 8.6\% and 21.3\% reductions in MSE and MAE, respectively, and further decreases MaxAE by 28\% compared with Transolver. Moreover, GA-Field delivers consistently smaller relative errors than TripNet, demonstrating superior accuracy. Qualitatively, Figure \ref{fig:surface_exp} shows that our predictions closely match the ground truth across the vehicle surface, including the complex underbody region. For 3D velocity, we observe consistent gains across all components. Compared with TripNet, our method reduces MSE, MAE, and relative errors by 8.9\%, 12.7\%, and 7.4\% / 16.7\% for the dominant component $U_x$, respectively. For the transverse components, MAE is reduced by 19.7\% on $U_y$ and 22.0\% on $U_z$. We also achieve consistent improvements for the magnitude $U$, with relative errors reduced by 7.1\% / 16.2\%. Figure \ref{fig:volume exp} further illustrates that our predictions capture key flow structures such as separation near the roof spoiler.

\begin{table}[!t]
  \caption{Detailed configurations of three out-of-distribution settings. Vehicles are categorized into three aerodynamic archetypes: Estateback, Notchback, and Fastback according to DrivAerNet++.}
  \label{tab: ood set}
  \setlength{\tabcolsep}{3pt}
  \begin{center}
    \begin{small}
      \begin{sc}
        \begin{tabular}{l|cc|cc}
          \toprule
                        
            \multirow{2}{*}{Settings} & \multicolumn{2}{c|}{Categories} & \multicolumn{2}{c}{Samples} \\
            \cmidrule(lr){2-3} \cmidrule(lr){4-5}
                                      & Train & Test & Train & Test\\
            \midrule

            \multirow{2}{*}{Setting 1} & Fastbacks   & \multirow{2}{*}{Estatebacks} & \multirow{2}{*}{5376} & \multirow{2}{*}{1386} \\
                                       & Notchbacks  &                              &                       & \\
            \midrule
            \multirow{2}{*}{Setting 2} & Fastbacks   & \multirow{2}{*}{Notchbacks}  & \multirow{2}{*}{5248} & \multirow{2}{*}{1403} \\
                                       & Estatebacks &                              &                       & \\
            \midrule
            \multirow{2}{*}{Setting 3} & Estatebacks & \multirow{2}{*}{Fastbacks}   & \multirow{2}{*}{2176} & \multirow{2}{*}{5332} \\
                                       & Notchbacks  &                              &                       & \\
            \bottomrule
        \end{tabular}
      \end{sc}
    \end{small}
  \end{center}
  \vskip -0.1in
\end{table}

\subsection{OOD Generalization}
\label{ood}
Out-of-distribution (OOD) generalization is a critical criterion for evaluating model robustness. Following CarBench \cite{elrefaie2025carbench}, we design three experimental settings to assess generalization across different vehicle categories. Detailed configurations of these settings are summarized in Table~\ref{tab: ood set}. During evaluation, we follow CarBench and report the relative $\ell_2$ error and coefficient of determination $R^2$. All metrics are computed on unnormalized pressure values using 10k points per test case. As shown in Table~\ref{tab: ood result}, our method achieves strong performance across all three settings. The advantage is particularly pronounced in Setting 3, which represents the most challenging generalization scenario. Compared with Transolver, our method improves relative $\ell_2$ error by 18.6\% and $R^2$ by 15.1\% under this setting. This improvement can be attributed to our design that promotes explicit geometry-flow coupling, which facilitates adaptation to substantial shape variations and improves generalization to unseen vehicle categories.

\begin{table}[!t]
  \caption{Performance comparison on the pressure field prediction task between Transolver \cite{wu2024Transolver}, Transolver++ \cite{luo2025transolver++}, and GA-Field under three OOD settings.}
  \label{tab: ood result}
  \setlength{\tabcolsep}{4pt}
  \begin{center}
    \begin{small}
      \begin{sc}
        \begin{tabular}{l|l|ccccc|c}
          \toprule
                        
            {Settings} & {Method} & {Rel L2$\downarrow$} &  {$R^2\uparrow$} \\
            \midrule
            \multirow{3}{*}{Setting 1} & Transolver \textsubscript{\tiny ICML\textquotesingle24}& 0.2391 & 0.9088 \\
                                       & Transolver++ \textsubscript{\tiny ICML\textquotesingle25}& \underline{0.2182} & \underline{0.9241} \\
                                       & GA-Field     & \textbf{0.2058} & \textbf{0.9302} \\
            \midrule
            \multirow{3}{*}{Setting 2} & Transolver \textsubscript{\tiny ICML\textquotesingle24}& \underline{0.1725} & \underline{0.9542} \\
                                       & Transolver++ \textsubscript{\tiny ICML\textquotesingle25}& 0.1729 & 0.9540 \\
                                       & GA-Field     & \textbf{0.1634} & \textbf{0.9568} \\
            \midrule
            \multirow{3}{*}{Setting 3} & Transolver \textsubscript{\tiny ICML\textquotesingle24}& \underline{0.4894} & \underline{0.6083} \\
                                       & Transolver++ \textsubscript{\tiny ICML\textquotesingle25}& {0.4988} & {0.5933} \\
                                       & GA-Field     & \textbf{0.3980} & \textbf{0.7004} \\
           
            \bottomrule
        \end{tabular}
      \end{sc}
    \end{small}
  \end{center}
  \vskip -0.1in
\end{table}

\subsection{Ablation Studies}
\label{ablation}
\noindent\textbf{Ablation of components.} We conduct ablation studies to evaluate the contribution of each component, including (i) grouped vector attention and grid pooling, (ii) task-specific input geometric features, (iii) the field detail refinement module, and (iv) the global geometry injection module. As shown in Table \ref{tab: ablation study}, introducing group vector attention and grid pooling reduces the relative errors by 5.9\% / 3.0\% over the baseline, validating the effectiveness of grid-based aggregation and grouped attention. On top of them, adding task-specific geometric cues such as \textit{inflow-angle} further reduces the errors by 4.3\% / 6.2\%. The refinement module yields a modest gain compared to other components, primarily because it predicts a residual correction whose magnitude is limited under the current regression objective. Finally, the global geometry injection module consistently improves performance, reducing both relative $\ell_1$ and relative $\ell_2$ errors by 3.9\%, which underscores the importance of global shape context for aerodynamic field prediction.

\noindent\textbf{Ablation of input scale.} 

Existing evaluations are typically conducted with a fixed number of sampled points \cite{chen2025tripnet,zou2026adafield}, which may not reveal the model’s sensitivity to input resolution at test time. To assess the robustness of GA-Field under different input scales, we keep the trained model unchanged and vary the number of sampled input points during inference. As shown in Table \ref{tab: evaluation scale}, the performance remains stable as the sampled points increase exponentially from 16,384 to 131,072, indicating strong scale stability of GA-Field. Meanwhile, MaxAE grows roughly linearly with the sampling size, primarily because of the increased chance of outliers when more points are included, making the worst-case error more prominent.

\begin{table}[!t]
  \caption{Ablation study of our method. The evaluation is conducted on the surface pressure field prediction task. GVA and GP represent the Grouped Vector Attention and Grid Pooling.}
  \label{tab: ablation study}
  \begin{center}
    \begin{small}
      \begin{sc}
        \begin{tabular}{l|cccc}
          \toprule

            {Method} & {Rel L2{(\%)$\downarrow$}} & {Rel L1{(\%)$\downarrow$}}\\
            \midrule
            {Baseline}                 & 22.56 & 19.58 \\
            {+ GVA \& GP}              & 21.24 & 18.98 \\
            {+ Geometric Feature}           & 20.31 & 17.80 \\
            {+ Detail Refinement}       & 20.08 & 17.70 \\
            {+ Geometry Injection}     & \textbf{19.28} & \textbf{17.00} \\
            \bottomrule
        \end{tabular}
      \end{sc}
    \end{small}
  \end{center}
  \vskip -0.1in
\end{table}

\begin{table}[!t]
  \caption{Ablation of input scale. The evaluation is conducted on the surface pressure field prediction task.}
  \label{tab: evaluation scale}
  \setlength{\tabcolsep}{3pt}
  \begin{center}
    \begin{small}
      \begin{sc}
        \begin{tabular}{l|ccccc}
          \toprule
            \multirow{2}{*}{Points} & {MSE$\downarrow$} & {MAE$\downarrow$} & \multirow{2}{*}{MaxAE$\downarrow$} & {Rel L2} & {Rel L1}\\
             
            & {($\times10^{-2}$)} & {($\times10^{-1}$)} &   & {(\%)$\downarrow$} & {(\%)$\downarrow$} \\
            \midrule
            {16384}      & 4.40 & 1.08 & 5.02 & 19.81 & 17.61 \\
            {32768}      & 4.13 & 1.04 & 6.13 & 19.28 & 17.00 \\
            {65536}      & 4.23 & 1.04 & 7.30 & 19.41 & 17.03 \\
            {131072}     & 4.56 & 1.07 & 8.54 & 19.95 & 17.52 \\
            \bottomrule
        \end{tabular}
      \end{sc}
    \end{small}
  \end{center}
  \vskip -0.1in
\end{table}

\begin{table}[!t]
  \caption{Performance comparison on the ShapeNet-Car benchmark. We report relative $\ell_2$ errors for surrounding (Volume) velocity and surface pressure.}
  \label{tab: shapenet}
  \begin{center}
      \begin{sc}
        \begin{tabular}{l|cc}
          \toprule
            {Method} & {Volume$\downarrow$} & {Surface$\downarrow$}  \\
            \midrule

            {Simple MLP}                                      & 0.0512 & 0.1304   \\
            {GraphSAGE \textsubscript{\tiny NIPS\textquotesingle17}}        & 0.0461 & 0.1050  \\
            {PointNet \textsubscript{\tiny CVPR\textquotesingle17}}                & 0.0494 & 0.1104   \\
            {Graph U-Net \textsubscript{\tiny ICML\textquotesingle19}}               & 0.0471 & 0.1102   \\
            {GNO \textsubscript{\tiny ICLRW\textquotesingle20}}               & 0.0383 & 0.0815   \\
            {MeshGraphNet \textsubscript{\tiny ICLR\textquotesingle21}}         & 0.0354 & 0.0781   \\
            {Galerkin \textsubscript{\tiny NIPS\textquotesingle21}}                 & 0.0339 & 0.0878   \\
            {geo-FNO \textsubscript{\tiny JMLR\textquotesingle23}}                  & 0.1670 & 0.2378   \\
            {GNOT \textsubscript{\tiny ICML\textquotesingle23}}                       & 0.0329 & 0.0798   \\
            {GINO \textsubscript{\tiny NIPS\textquotesingle23}}              & 0.0386 & 0.0810   \\
            {3D-GeoCA \textsubscript{\tiny IJCAI\textquotesingle24}}                 & 0.0319 & 0.0779   \\
            {Transolver \textsubscript{\tiny ICML\textquotesingle24}}            & 0.0228 & 0.0793   \\
            {SpiderSolver \textsubscript{\tiny NIPS\textquotesingle25}}        & \underline{0.0210} & \underline{0.0738}   \\

            \midrule
            {GA-Field}                                        & \textbf{0.0206} & \textbf{0.0726}   \\
            \bottomrule
        \end{tabular}
      \end{sc}
  \end{center}
  \vskip -0.1in
\end{table}

\subsection{ShapeNet-Car}
\label{shapenetcar}
To verify the effectiveness of our method beyond DrivAerNet++, we conduct comparative experiments on an additional well-established automotive CFD benchmark, ShapeNet-Car \cite{shapnetcar}. In this work, we focus on the aerodynamic field task, and we report the relative $\ell_2$ error for surface pressure and surrounding velocity, following the evaluation protocol in \cite{wu2024Transolver, qi2025spidersolver}. We compare GA-Field with the various baselines including GraphSAGE \cite{hamilton2017inductive}, PointNet \cite{qi2017pointnet}, Graph U-Net \cite{gao2019graph}, MeshGraphNet \cite{pfaff2021learning}, GNO \cite{anandkumar2019neural}, Galerkin \cite{cao2021choose}, geo-FNO \cite{li2023fourier}, GNOT \cite{hao2023gnot}, GINO \cite{NEURIPS2023_70518ea4}, 3D-GeoCA \cite{ijcai2024p640}, Transolver \cite{wu2024Transolver}, and SpiderSolver \cite{qi2025spidersolver}. As shown in Table \ref{tab: shapenet}, GA-Field achieves state-of-the-art performance on ShapeNet-Car, outperforming a wide range of strong baselines, highlighting the strong generalization ability of GA-Field across diverse vehicle datasets.

\subsection{Analysis of Part-Wise Drag Contribution}
\label{part drag}

\begin{figure}[t]
    \centering
    \begin{minipage}{\linewidth}
        \centering
		\includegraphics[width=\linewidth]{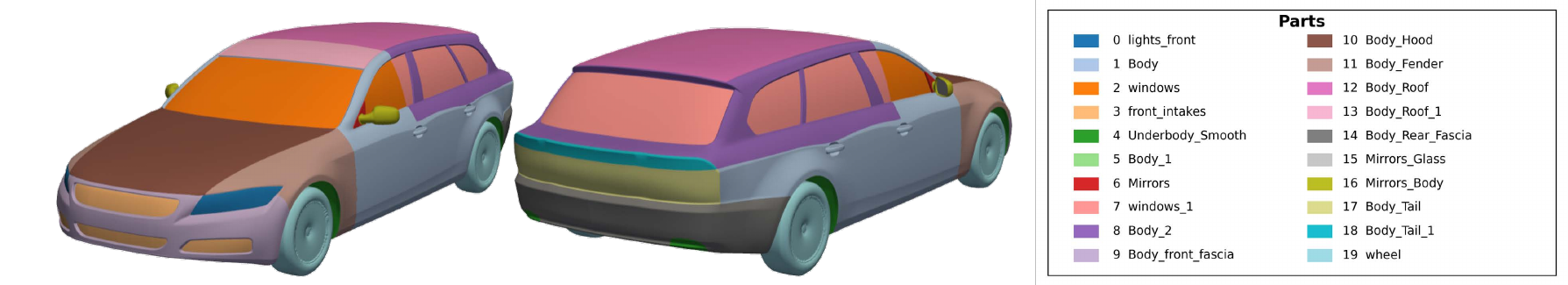}
        \centerline{\footnotesize{ (a) Part-wise annotations.}}
	\end{minipage}
    
    \begin{minipage}{\linewidth}
        \centering
		\includegraphics[width=\linewidth]{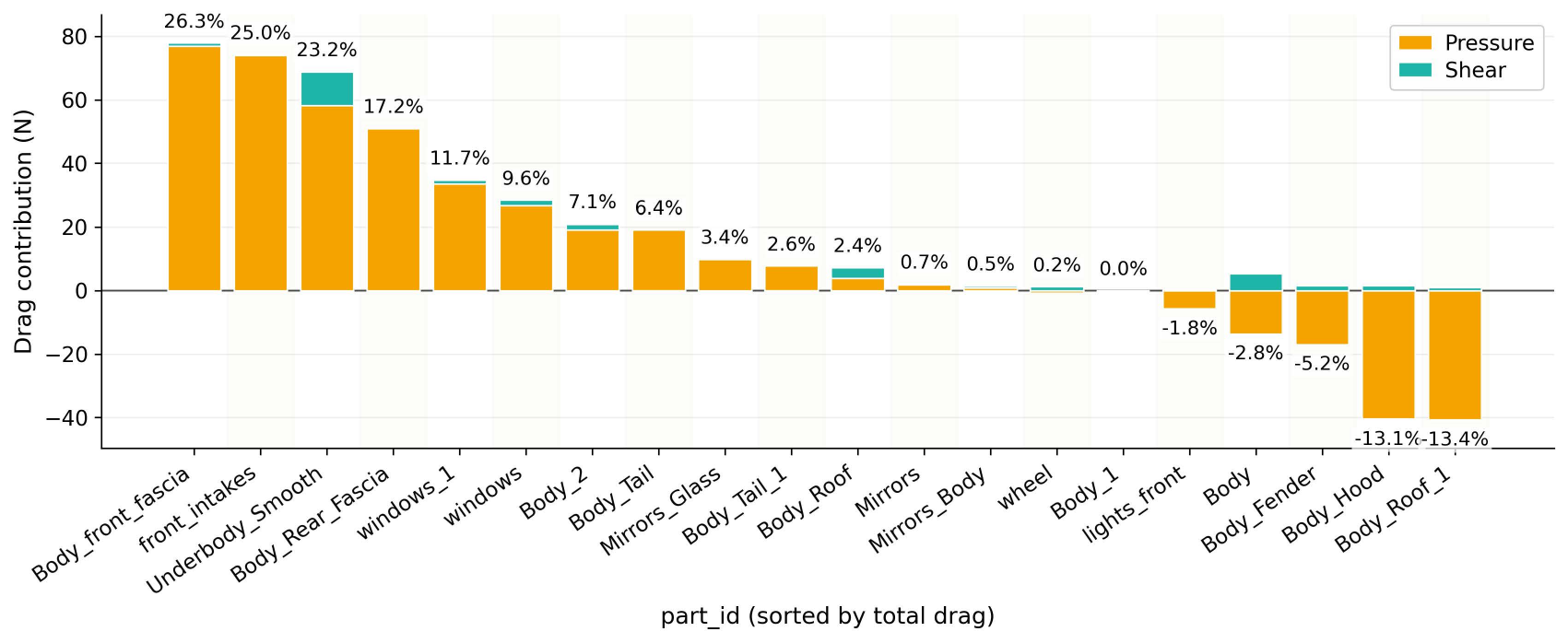}
		\centerline{\footnotesize{ (b) Drag force contribution of each part.}}
	\end{minipage}
    \caption{Part-wise annotation visualization and part-wise drag-force contribution for car design E\_S\_WWC\_WM\_094. The car is annotated with 20 part labels. The y-axis in (b) indicates the drag force on each component, measured in newtons.}
    \label{fig:drag force}
\end{figure}

Typical drag-coefficient prediction methods \cite{wang2025aerodynamic,chen2025tripnet,choy2025factorized} learns to map vehicle geometry to a single scalar, thereby providing limited guidance for part-level design optimization. In contrast, we compute part-wise drag for each annotated region, as shown in Figure \ref{fig:drag force}, which provides a detailed breakdown of region-specific drag mechanisms. It can be observed that most parts are pressure-dominated, with the front fascia and intakes together accounting for about 51\% of the total drag, suggesting that reducing front-end pressure drag is the main optimization target. The underbody contributes approximately 23\% but exhibits the largest relative shear share, motivating optimization strategies targeting near-wall shear. The rear fascia remains a non-negligible source at around 17\%, again dominated by pressure effects. Interestingly, several regions show negative contributions, such as the hood and roof at approximately -13\% each, implying favorable pressure effects that should be preserved. Overall, effective drag reduction should combine pressure-focused improvements on the front-end and rear with shear-focused mitigation on the underbody, while maintaining the beneficial load distribution observed on the hood and roof.

\section{Conclusion and Future Work}

 In this work, we presented GA-Field, a geometry-aware neural network for vehicle aerodynamic field prediction. By injecting global geometry embeddings at multiple network stages and introducing a coarse-to-fine detail refinement strategy, GA-Field preserves long-range shape awareness while effectively correcting fine-scale errors. Extensive experiments on ShapeNet-Car and DrivAerNet++ benchmark show that GA-Field achieves state-of-the-art performance on multiple field prediction tasks, while exhibiting strong out-of-distribution generalization across different vehicle categories. These results underscore the importance of repeated geometric conditioning and multi-scale refinement in aerodynamic field learning, and suggest a promising direction for efficient, geometry-aware modeling. In the future, we plan to incorporate physics-based constraints into GA-Field to improve physical consistency and enable more physically meaningful aerodynamic field prediction.

\section*{Impact Statement}

GA-Field accelerates vehicle aerodynamic evaluation by predicting high-fidelity flow fields directly from geometry and flow conditions, enabling fast screening and rapid design iteration when full CFD is the main bottleneck. By predicting surface pressure, wall shear stress, and extending to 3D velocity, GA-Field supports physics-based analysis beyond scalar coefficient prediction, including part-wise load and drag estimation that helps engineers identify which components contribute most. This can lower the barrier to aerodynamic assessment under limited compute budgets and reduce overall computation by cutting down the number of expensive CFD runs needed during exploration. For safety-critical decisions, GA-Field should be used as a decision support tool together with established verification methods such as CFD or wind tunnel testing.


\bibliography{ref}

@misc{chen2025tripnet,
      title={TripNet: Learning Large-scale High-fidelity 3D Car Aerodynamics with Triplane Networks}, 
      author={Qian Chen and Mohamed Elrefaie and Angela Dai and Faez Ahmed},
      year={2025},
      eprint={2503.17400},
      archivePrefix={arXiv},
      primaryClass={physics.flu-dyn},
      url={https://arxiv.org/abs/2503.17400}, 
}

@InProceedings{wu2024Transolver,
  title = 	 {Transolver: A Fast Transformer Solver for {PDE}s on General Geometries},
  author =       {Wu, Haixu and Luo, Huakun and Wang, Haowen and Wang, Jianmin and Long, Mingsheng},
  booktitle = 	 {Proceedings of the 41st International Conference on Machine Learning},
  pages = 	 {53681--53705},
  year = 	 {2024},
  editor = 	 {Salakhutdinov, Ruslan and Kolter, Zico and Heller, Katherine and Weller, Adrian and Oliver, Nuria and Scarlett, Jonathan and Berkenkamp, Felix},
  volume = 	 {235},
  series = 	 {Proceedings of Machine Learning Research},
  month = 	 {21--27 Jul},
  publisher =    {PMLR},
  url = 	 {https://proceedings.mlr.press/v235/wu24r.html},
}

@InProceedings{ronneberger2015u,
author="Ronneberger, Olaf
and Fischer, Philipp
and Brox, Thomas",
editor="Navab, Nassir
and Hornegger, Joachim
and Wells, William M.
and Frangi, Alejandro F.",
title="U-Net: Convolutional Networks for Biomedical Image Segmentation",
booktitle="Medical Image Computing and Computer-Assisted Intervention -- MICCAI 2015",
year="2015",
publisher="Springer International Publishing",
address="Cham",
pages="234--241",
}

@inproceedings{wu2022point,
 author = {Wu, Xiaoyang and Lao, Yixing and Jiang, Li and Liu, Xihui and Zhao, Hengshuang},
 booktitle = {Advances in Neural Information Processing Systems},
 editor = {S. Koyejo and S. Mohamed and A. Agarwal and D. Belgrave and K. Cho and A. Oh},
 pages = {33330--33342},
 publisher = {Curran Associates, Inc.},
 title = {Point Transformer V2: Grouped Vector Attention and Partition-based Pooling},
 url = {https://proceedings.neurips.cc/paper_files/paper/2022/file/d78ece6613953f46501b958b7bb4582f-Paper-Conference.pdf},
 volume = {35},
 year = {2022}
}

@ARTICLE{eldar1997farthest,
  author={Eldar, Y. and Lindenbaum, M. and Porat, M. and Zeevi, Y.Y.},
  journal={IEEE Transactions on Image Processing}, 
  title={The farthest point strategy for progressive image sampling}, 
  year={1997},
  volume={6},
  number={9},
  pages={1305-1315},
  doi={10.1109/83.623193}}

@misc{choy2025factorized,
      title={Factorized Implicit Global Convolution for Automotive Computational Fluid Dynamics Prediction}, 
      author={Chris Choy and Alexey Kamenev and Jean Kossaifi and Max Rietmann and Jan Kautz and Kamyar Azizzadenesheli},
      year={2025},
      eprint={2502.04317},
      archivePrefix={arXiv},
      primaryClass={cs.CV},
      url={https://arxiv.org/abs/2502.04317}, 
}

@inproceedings{elrefaie2024drivaernet++,
 author = {Elrefaie, Mohamed and Morar, Florin and Dai, Angela and Ahmed, Faez},
 booktitle = {Advances in Neural Information Processing Systems},
 doi = {10.52202/079017-0016},
 editor = {A. Globerson and L. Mackey and D. Belgrave and A. Fan and U. Paquet and J. Tomczak and C. Zhang},
 pages = {499--536},
 publisher = {Curran Associates, Inc.},
 title = {DrivAerNet++: A Large-Scale Multimodal Car Dataset with Computational Fluid Dynamics Simulations and Deep Learning Benchmarks},
 url = {https://proceedings.neurips.cc/paper_files/paper/2024/file/013cf29a9e68e4411d0593040a8a1eb3-Paper-Datasets_and_Benchmarks_Track.pdf},
 volume = {37},
 year = {2024}
}

@article{perez2018film, title={FiLM: Visual Reasoning with a General Conditioning Layer}, volume={32}, url={https://ojs.aaai.org/index.php/AAAI/article/view/11671}, DOI={10.1609/aaai.v32i1.11671}, number={1}, journal={Proceedings of the AAAI Conference on Artificial Intelligence}, author={Perez, Ethan and Strub, Florian and de Vries, Harm and Dumoulin, Vincent and Courville, Aaron}, year={2018}, month={Apr.} }

@inproceedings{zou2026adafield,
  title        = {AdaField: Generalizable Surface Pressure Modeling with Physics-Informed Pre-training and Flow-Conditioned Adaptation},
  author       = {Zou, Junhong and Qiu, Wei and Sun, Zhenxu and Zhang, Xiaomei and Zhang, Zhaoxiang and Zhu, Xiangyu},
  booktitle    = {Proceedings of the AAAI Conference on Artificial Intelligence},
  year         = {2026}
}

@article{hosseini2011pressure,
title = {Pressure boundary conditions for computing incompressible flows with SPH},
journal = {Journal of Computational Physics},
volume = {230},
number = {19},
pages = {7473-7487},
year = {2011},
issn = {0021-9991},
doi = {https://doi.org/10.1016/j.jcp.2011.06.013},
url = {https://www.sciencedirect.com/science/article/pii/S0021999111003779},
author = {S. Majid Hosseini and James J. Feng},
}

@article{vreman2014projection,
title = {The projection method for the incompressible Navier–Stokes equations: The pressure near a no-slip wall},
journal = {Journal of Computational Physics},
volume = {263},
pages = {353-374},
year = {2014},
issn = {0021-9991},
doi = {https://doi.org/10.1016/j.jcp.2014.01.035},
url = {https://www.sciencedirect.com/science/article/pii/S002199911400062X},
author = {A.W. Vreman},
}

@inproceedings{elrefaie2024drivaernet,
  title={Drivaernet: A parametric car dataset for data-driven aerodynamic design and graph-based drag prediction},
  author={Elrefaie, Mohamed and Dai, Angela and Ahmed, Faez},
  booktitle={International Design Engineering Technical Conferences and Computers and Information in Engineering Conference},
  volume={88360},
  pages={V03AT03A019},
  year={2024},
  organization={American Society of Mechanical Engineers},
  doi = {10.1115/DETC2024-143593},
  url = {https://doi.org/10.1115/DETC2024-143593},
}

@article{wang2025aerodynamic, title={Aerodynamic Coefficients Prediction via Cross-Attention Fusion and Physical-Informed Training}, volume={39}, url={https://ojs.aaai.org/index.php/AAAI/article/view/32071}, DOI={10.1609/aaai.v39i1.32071}, number={1}, journal={Proceedings of the AAAI Conference on Artificial Intelligence}, author={Wang, Yueqing and Zhang, Peng and Liu, Yushuang and Zhao, Jianing and Lin, Jie and Chen, Yi}, year={2025}, month={Apr.}, pages={869-876} }

@article{gao2025accurate,
author = {Gao, Hongrui and Liu, Tanghong and Chen, Xiaodong and Huo, Xiaoshuai and Chen, Zhengwei and Zhang, Jie and Khoo, Boo},
year = {2025},
month = {02},
pages = {1423-1453},
title = {An accurate and efficient methodology on wind spectra relative to moving trains: field measurements of wind characteristics in complex terrains},
volume = {39},
journal = {Stochastic Environmental Research and Risk Assessment},
doi = {10.1007/s00477-025-02924-2}
}

@inproceedings{qi2017pointnet,
  title={Pointnet: Deep learning on point sets for 3d classification and segmentation},
  author={Qi, Charles R and Su, Hao and Mo, Kaichun and Guibas, Leonidas J},
  booktitle={Proceedings of the IEEE conference on computer vision and pattern recognition},
  pages={652--660},
  year={2017}
}

@inproceedings{zhao2021point,
  title={Point transformer},
  author={Zhao, Hengshuang and Jiang, Li and Jia, Jiaya and Torr, Philip HS and Koltun, Vladlen},
  booktitle={Proceedings of the IEEE/CVF international conference on computer vision},
  pages={16259--16268},
  year={2021}
}

@inproceedings{
luo2025transolver++,
title={Transolver++: An Accurate Neural Solver for {PDE}s on Million-Scale Geometries},
author={Huakun Luo and Haixu Wu and Hang Zhou and Lanxiang Xing and Yichen Di and Jianmin Wang and Mingsheng Long},
booktitle={Forty-second International Conference on Machine Learning},
year={2025},
url={https://openreview.net/forum?id=AM7iAh0krx}
}

@inproceedings{
li2021fourier,
title={Fourier Neural Operator for Parametric Partial Differential Equations},
author={Zongyi Li and Nikola Borislavov Kovachki and Kamyar Azizzadenesheli and Burigede liu and Kaushik Bhattacharya and Andrew Stuart and Anima Anandkumar},
booktitle={International Conference on Learning Representations},
year={2021},
url={https://openreview.net/forum?id=c8P9NQVtmnO}
}

@inproceedings{
pfaff2021learning,
title={Learning Mesh-Based Simulation with Graph Networks},
author={Tobias Pfaff and Meire Fortunato and Alvaro Sanchez-Gonzalez and Peter Battaglia},
booktitle={International Conference on Learning Representations},
year={2021},
url={https://openreview.net/forum?id=roNqYL0_XP}
}

@misc{nabian2024x,
      title={X-MeshGraphNet: Scalable Multi-Scale Graph Neural Networks for Physics Simulation}, 
      author={Mohammad Amin Nabian and Chang Liu and Rishikesh Ranade and Sanjay Choudhry},
      year={2024},
      eprint={2411.17164},
      archivePrefix={arXiv},
      primaryClass={cs.LG},
      url={https://arxiv.org/abs/2411.17164}, 
}

@article{pang2023masked,
  title={Masked autoencoders for 3d point cloud self-supervised learning},
  author={Pang, Yatian and Tay, Eng Hock Francis and Yuan, Li and Chen, Zhenghua},
  journal={World Scientific Annual Review of Artificial Intelligence},
  volume={1},
  pages={2440001},
  year={2023},
  publisher={World Scientific}
}

@misc{elrefaie2025carbench,
      title={CarBench: A Comprehensive Benchmark for Neural Surrogates on High-Fidelity 3D Car Aerodynamics}, 
      author={Mohamed Elrefaie and Dule Shu and Matt Klenk and Faez Ahmed},
      year={2025},
      eprint={2512.07847},
      archivePrefix={arXiv},
      primaryClass={cs.LG},
      url={https://arxiv.org/abs/2512.07847}, 
}

@inproceedings{NEURIPS2023_70518ea4,
 author = {Li, Zongyi and Kovachki, Nikola and Choy, Chris and Li, Boyi and Kossaifi, Jean and Otta, Shourya and Nabian, Mohammad Amin and Stadler, Maximilian and Hundt, Christian and Azizzadenesheli, Kamyar and Anandkumar, Animashree},
 booktitle = {Advances in Neural Information Processing Systems},
 editor = {A. Oh and T. Naumann and A. Globerson and K. Saenko and M. Hardt and S. Levine},
 pages = {35836--35854},
 publisher = {Curran Associates, Inc.},
 title = {Geometry-Informed Neural Operator for Large-Scale 3D PDEs},
 url = {https://proceedings.neurips.cc/paper_files/paper/2023/file/70518ea42831f02afc3a2828993935ad-Paper-Conference.pdf},
 volume = {36},
 year = {2023}
}

@inproceedings{ijcai2024p640,
  title     = {Geometry-Guided Conditional Adaptation for Surrogate Models of Large-Scale 3D PDEs on Arbitrary Geometries},
  author    = {Deng, Jingyang and Li, Xingjian and Xiong, Haoyi and Hu, Xiaoguang and Ma, Jinwen},
  booktitle = {Proceedings of the Thirty-Third International Joint Conference on
               Artificial Intelligence, {IJCAI-24}},
  publisher = {International Joint Conferences on Artificial Intelligence Organization},
  editor    = {Kate Larson},
  pages     = {5790--5798},
  year      = {2024},
  month     = {8},
  note      = {Main Track},
  doi       = {10.24963/ijcai.2024/640},
  url       = {https://doi.org/10.24963/ijcai.2024/640}
}

@inproceedings{
qi2025spidersolver,
title={SpiderSolver: A Geometry-Aware Transformer for Solving {PDE}s on Complex Geometries},
author={Kai Qi and Fan Wang and Zhewen Dong and Jian Sun},
booktitle={The Thirty-ninth Annual Conference on Neural Information Processing Systems},
year={2025},
url={https://openreview.net/forum?id=hWtvsL51hO}
}

@inproceedings{hamilton2017inductive,
 author = {Hamilton, Will and Ying, Zhitao and Leskovec, Jure},
 booktitle = {Advances in Neural Information Processing Systems},
 editor = {I. Guyon and U. Von Luxburg and S. Bengio and H. Wallach and R. Fergus and S. Vishwanathan and R. Garnett},
 pages = {},
 publisher = {Curran Associates, Inc.},
 title = {Inductive Representation Learning on Large Graphs},
 url = {https://proceedings.neurips.cc/paper_files/paper/2017/file/5dd9db5e033da9c6fb5ba83c7a7ebea9-Paper.pdf},
 volume = {30},
 year = {2017}
}

@InProceedings{gao2019graph,
  title = 	 {Graph U-Nets},
  author =       {Gao, Hongyang and Ji, Shuiwang},
  booktitle = 	 {Proceedings of the 36th International Conference on Machine Learning},
  pages = 	 {2083--2092},
  year = 	 {2019},
  editor = 	 {Chaudhuri, Kamalika and Salakhutdinov, Ruslan},
  volume = 	 {97},
  series = 	 {Proceedings of Machine Learning Research},
  month = 	 {09--15 Jun},
  publisher =    {PMLR},
  url = 	 {https://proceedings.mlr.press/v97/gao19a.html},
}

@inproceedings{
anandkumar2019neural,
title={Neural Operator: Graph Kernel Network for Partial Differential Equations},
author={Anima Anandkumar and Kamyar Azizzadenesheli and Kaushik Bhattacharya and Nikola Kovachki and Zongyi Li and Burigede Liu and Andrew Stuart},
booktitle={ICLR 2020 Workshop on Integration of Deep Neural Models and Differential Equations},
year={2019},
url={https://openreview.net/forum?id=fg2ZFmXFO3}
}

@inproceedings{cao2021choose,
 author = {Cao, Shuhao},
 booktitle = {Advances in Neural Information Processing Systems},
 editor = {M. Ranzato and A. Beygelzimer and Y. Dauphin and P.S. Liang and J. Wortman Vaughan},
 pages = {24924--24940},
 publisher = {Curran Associates, Inc.},
 title = {Choose a Transformer: Fourier or Galerkin},
 url = {https://proceedings.neurips.cc/paper_files/paper/2021/file/d0921d442ee91b896ad95059d13df618-Paper.pdf},
 volume = {34},
 year = {2021}
}

@article{li2023fourier,
  author  = {Zongyi Li and Daniel Zhengyu Huang and Burigede Liu and Anima Anandkumar},
  title   = {Fourier Neural Operator with Learned Deformations for PDEs on General Geometries},
  journal = {Journal of Machine Learning Research},
  year    = {2023},
  volume  = {24},
  number  = {388},
  pages   = {1--26},
  url     = {http://jmlr.org/papers/v24/23-0064.html}
}

@InProceedings{hao2023gnot,
  title = 	 {{GNOT}: A General Neural Operator Transformer for Operator Learning},
  author =       {Hao, Zhongkai and Wang, Zhengyi and Su, Hang and Ying, Chengyang and Dong, Yinpeng and Liu, Songming and Cheng, Ze and Song, Jian and Zhu, Jun},
  booktitle = 	 {Proceedings of the 40th International Conference on Machine Learning},
  pages = 	 {12556--12569},
  year = 	 {2023},
  editor = 	 {Krause, Andreas and Brunskill, Emma and Cho, Kyunghyun and Engelhardt, Barbara and Sabato, Sivan and Scarlett, Jonathan},
  volume = 	 {202},
  series = 	 {Proceedings of Machine Learning Research},
  month = 	 {23--29 Jul},
  publisher =    {PMLR},
  url = 	 {https://proceedings.mlr.press/v202/hao23c.html},
}

@article{shapnetcar,
author = {Umetani, Nobuyuki and Bickel, Bernd},
title = {Learning three-dimensional flow for interactive aerodynamic design},
year = {2018},
issue_date = {August 2018},
publisher = {Association for Computing Machinery},
address = {New York, NY, USA},
volume = {37},
number = {4},
issn = {0730-0301},
url = {https://doi.org/10.1145/3197517.3201325},
doi = {10.1145/3197517.3201325},
journal = {ACM Trans. Graph.},
month = jul,
articleno = {89},
numpages = {10}
}

@misc{chang2015shapenet,
      title={ShapeNet: An Information-Rich 3D Model Repository}, 
      author={Angel X. Chang and Thomas Funkhouser and Leonidas Guibas and Pat Hanrahan and Qixing Huang and Zimo Li and Silvio Savarese and Manolis Savva and Shuran Song and Hao Su and Jianxiong Xiao and Li Yi and Fisher Yu},
      year={2015},
      eprint={1512.03012},
      archivePrefix={arXiv},
      primaryClass={cs.GR},
      url={https://arxiv.org/abs/1512.03012}, 
}

@inproceedings{
loshchilov2018decoupled,
title={Decoupled Weight Decay Regularization},
author={Ilya Loshchilov and Frank Hutter},
booktitle={International Conference on Learning Representations},
year={2019},
url={https://openreview.net/forum?id=Bkg6RiCqY7},
}

@InProceedings{Du_2025_CVPR,
    author    = {Du, Yi and Zhao, Zhipeng and Su, Shaoshu and Golluri, Sharath and Zheng, Haoze and Yao, Runmao and Wang, Chen},
    title     = {SuperPC: A Single Diffusion Model for Point Cloud Completion, Upsampling, Denoising, and Colorization},
    booktitle = {Proceedings of the IEEE/CVF Conference on Computer Vision and Pattern Recognition (CVPR)},
    month     = {June},
    year      = {2025},
    pages     = {16953-16964}
}

@InProceedings{Huang_2021_ICCV,
    author    = {Huang, Hsin-Ping and Tseng, Hung-Yu and Saini, Saurabh and Singh, Maneesh and Yang, Ming-Hsuan},
    title     = {Learning To Stylize Novel Views},
    booktitle = {Proceedings of the IEEE/CVF International Conference on Computer Vision (ICCV)},
    month     = {October},
    year      = {2021},
    pages     = {13869-13878}
}

@misc{jun2023shapegeneratingconditional3d,
      title={Shap-E: Generating Conditional 3D Implicit Functions}, 
      author={Heewoo Jun and Alex Nichol},
      year={2023},
      eprint={2305.02463},
      archivePrefix={arXiv},
      primaryClass={cs.CV},
      url={https://arxiv.org/abs/2305.02463}, 
}

@article{
rebain2023attention,
title={Attention Beats Concatenation for Conditioning Neural Fields},
author={Daniel Rebain and Mark J. Matthews and Kwang Moo Yi and Gopal Sharma and Dmitry Lagun and Andrea Tagliasacchi},
journal={Transactions on Machine Learning Research},
issn={2835-8856},
year={2023},
url={https://openreview.net/forum?id=GzqdMrFQsE},
note={}
}

@article{URQUHART2018308,
title = {Numerical analysis of a vehicle wake with tapered rear extensions under yaw conditions},
journal = {Journal of Wind Engineering and Industrial Aerodynamics},
volume = {179},
pages = {308-318},
year = {2018},
issn = {0167-6105},
doi = {https://doi.org/10.1016/j.jweia.2018.06.001},
url = {https://www.sciencedirect.com/science/article/pii/S0167610518301144},
author = {Magnus Urquhart and Simone Sebben and Lennert Sterken},
}

@article{WALLACE2025106177,
title = {Investigation of the link between vehicle underbody and base unsteady wake aerodynamics},
journal = {Journal of Wind Engineering and Industrial Aerodynamics},
volume = {265},
pages = {106177},
year = {2025},
issn = {0167-6105},
doi = {https://doi.org/10.1016/j.jweia.2025.106177},
url = {https://www.sciencedirect.com/science/article/pii/S0167610525001734},
author = {C. Wallace and A. Garmory and A. Gaylard and D. Butcher},
}

@article{WESSELING2001311,
title = {Geometric multigrid with applications to computational fluid dynamics},
journal = {Journal of Computational and Applied Mathematics},
volume = {128},
number = {1},
pages = {311-334},
year = {2001},
note = {Numerical Analysis 2000. Vol. VII: Partial Differential Equations},
issn = {0377-0427},
doi = {https://doi.org/10.1016/S0377-0427(00)00517-3},
url = {https://www.sciencedirect.com/science/article/pii/S0377042700005173},
author = {P. Wesseling and C.W. Oosterlee},
}

@article{YAN2007445,
title = {A modified full multigrid algorithm for the Navier–Stokes equations},
journal = {Computers \& Fluids},
volume = {36},
number = {2},
pages = {445-454},
year = {2007},
issn = {0045-7930},
doi = {https://doi.org/10.1016/j.compfluid.2006.01.017},
url = {https://www.sciencedirect.com/science/article/pii/S0045793006000430},
author = {J. Yan and F. Thiele and L. Xue},
}

@inproceedings{jayasankar2022defect,
author = {Akhil Jayasankar and Carl F. Ollivier Gooch},
title = {Defect Correction on Unstructured Finite Volume Solvers},
booktitle = {AIAA SCITECH 2022 Forum},
chapter = {},
pages={0220},
year={2022},
doi = {10.2514/6.2022-0220},
URL = {https://arc.aiaa.org/doi/abs/10.2514/6.2022-0220},
eprint = {https://arc.aiaa.org/doi/pdf/10.2514/6.2022-0220},
}

@inproceedings{elrefaie2024surrogate,
author = {Mohamed Elrefaie and Tarek Ayman and Mayar Elrefaie and Eman Sayed and Mahmoud Ayyad and Mohamed M. AbdelRahman},
title = {Surrogate Modeling of the Aerodynamic Performance for Airfoils in Transonic Regime},
booktitle = {AIAA SCITECH 2024 Forum},
chapter = {},
pages={2220},
year={2024},
doi = {10.2514/6.2024-2220},
URL = {https://arc.aiaa.org/doi/abs/10.2514/6.2024-2220},
eprint = {https://arc.aiaa.org/doi/pdf/10.2514/6.2024-2220},
}
\bibliographystyle{icml2026}

\newpage
\appendix

\section{Additional Experiment Results}

\begin{table}[!b]
  \caption{Drag force decomposition of each part in car design E\_S\_WWC\_WM\_094.}
  \label{tab: part drag}
  \begin{center}
    \begin{small}
      \begin{sc}
        \begin{tabular}{l|cc|c}
          \toprule
            \multirow{2}{*}{Part} & \multicolumn{2}{c|}{Drag Force ($N$)} &  {Area}\\
            \cmidrule(lr){2-3}
             & {$F_p$} & {$F_\tau$} & ($m^2$)\\
            \midrule
            {Body\_front\_fascia}                & 76.862 &  1.046  &  0.855 \\
            {front\_intakes}                     & 74.081 &  0.090  &  0.245 \\
            {Underbody\_Smooth}                  & 58.156 & 10.627  & 10.473 \\
            {Body\_Rear\_Fascia}                 & 50.924 &  0.106  &  0.979 \\
            {windows\_1}                         & 33.583 &  1.118  &  1.608 \\
            {windows}                            & 26.692 &  1.685  &  1.502 \\
            {Body\_2}                            & 19.121 &  1.798  &  1.769 \\
            {Body\_Tail}                         & 19.105 & -0.008  & 0.373 \\
            {Mirrors\_Glass}                     & 9.924 &  0.006  &  0.037 \\
            {Body\_Tail\_1}                      & 7.764 &  -0.004  &  0.152 \\
            {Body\_Roof}                         & 3.969 &  3.250  &  2.208 \\
            {Mirrors}                            & 1.943 &  0.038  &  0.029 \\
            {Mirrors\_Body}                      & 0.856 &  0.552  &  0.160 \\
            {wheel}                              & -0.699 &  1.253  &  4.366 \\
            {Body\_1}                            & 0.001 &  0.001  &  0.001 \\
            {lights\_front}                      & -5.725 &  0.289  &  0.139 \\
            {Body}                               & -13.614 &  5.288  &  3.855 \\
            {Body\_Fender}                       & -16.94 &  1.579  &  1.052 \\
            {Body\_Hood}                         &-40.430 &  1.624  &  1.848 \\
            {front\_Roof\_1}                     &-40.689 &  1.026  &  0.535 \\
            \bottomrule
        \end{tabular}
      \end{sc}
    \end{small}
  \end{center}
  \vskip -0.1in
\end{table}

\subsection{Analysis of Part-Wise Drag Contribution}
Previous drag-coefficient prediction methods \cite{wang2025aerodynamic,chen2025tripnet,choy2025factorized} output a single scalar, which provides limited guidance for part-level design optimization. Using part annotations and our predicted pressure and WSS fields, we are able to compute part-wise drag by aggregating cell-wise surface forces over each annotated region. Following \cite{chen2025tripnet}, the drag force of surface cell $i$ is: 
\begin{equation}
\begin{aligned}
        F_i = \underbrace{-\mathbf{n}_{i} \cdot p_{i}\cdot \mathbf{d}_{\infty} \cdot A_{i} \cdot \rho}_{F_{p,i}} + \underbrace{ \tau_{i} \cdot \mathbf{d}_{\infty} \cdot A_{i} \cdot \rho}_{F_{\tau,i}}
\end{aligned}
\end{equation}
where $F_i$ is the total drag force, $F_{p,i}$ and $F_{\tau,i}$ are drag force induced by pressure and WSS, $\mathbf{n}_i$ is the cell normal, $A_i$ is the cell area, and $\rho$ is the density of air. Detailed drag force decomposition is listed in Table \ref{tab: part drag}.

\subsection{Time and Memory Usage}
\begin{figure}[t]
    \centering
    \begin{minipage}{\linewidth}
        \centering
		\includegraphics[width=0.95\linewidth]{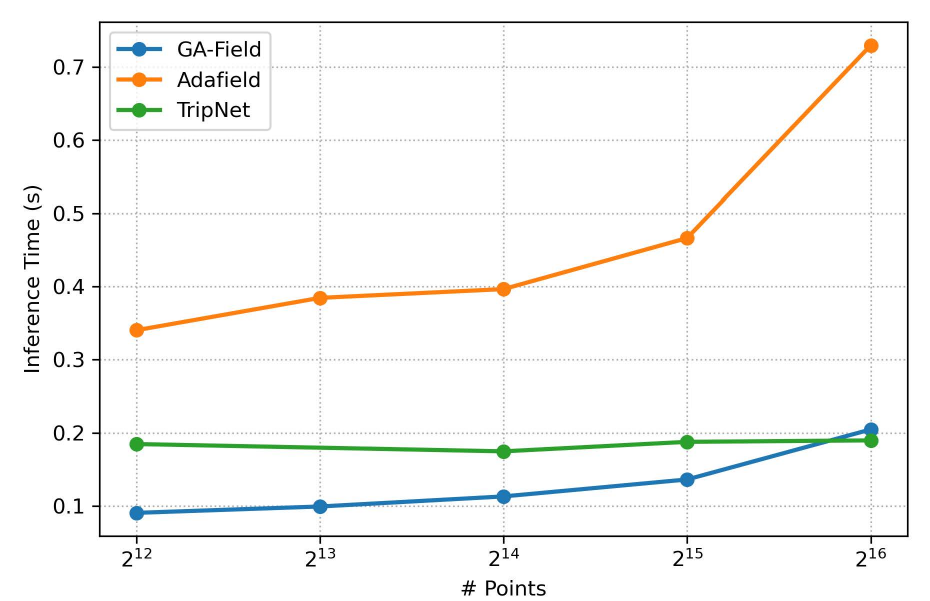}
        \centerline{\footnotesize{ (a) Inference Time vs. Number of Points}}
	\end{minipage}

    \begin{minipage}{\linewidth}
        \centering
		\includegraphics[width=0.95\linewidth]{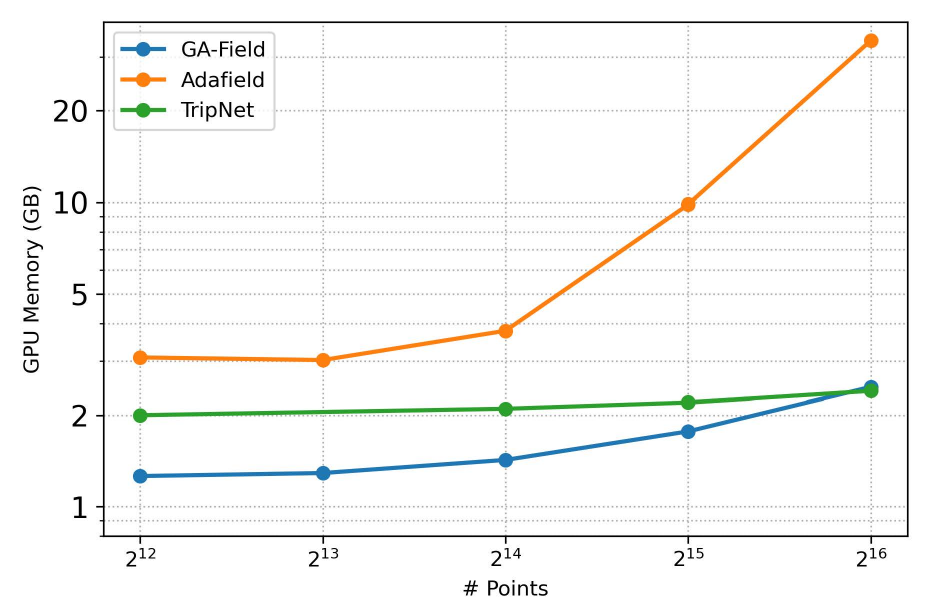}
		\centerline{\footnotesize{ (b) GPU Memory Usage vs. Number of Points}}
	\end{minipage}
    \caption{Scalability comparison of TripNet, AdaField, and GA-Field. }
    \label{fig:time_mem}
\end{figure}

We evaluate inference latency and peak GPU memory of GA-Field across input scales, and compare with TripNet \cite{chen2025tripnet} and AdaField \cite{zou2026adafield} (Fig.~\ref{fig:time_mem}). When the number of input points is below $2^{16}$, GA-Field is consistently faster and more memory-efficient than the triplane-based TripNet and the point-based AdaField. At $2^{16}$ points, GA-Field exhibits a similar latency and memory footprint to TripNet, indicating that our approach offers a clear efficiency advantage for small to medium input sizes.

\section{Implementation Details}
\label{add imp}
\subsection{Datasets}

\textbf{DrivAerNet++.} DrivAerNet++ \cite{elrefaie2024drivaernet++} is a large-scale benchmark for learning-based aerodynamic analysis of vehicles. It contains 8,121 diverse car designs, spanning fastback, notchback, and estateback configurations with varied wheel and underbody setups, each paired with high-fidelity CFD simulations and multiple data modalities (e.g., 3D meshes, surface point clouds, parametric models, CFD outputs, and part annotations). Beyond global coefficients (e.g., $C_d$, $C_l$), the dataset provides rich physical supervision, including velocity ($u$), pressure ($p$), and wall shear stress ($\tau$) fields. Each design is simulated under a uniform inlet wind speed of 30 m/s (108 km/h) in steady state RANS CFD. In our experiments, we use the official data split released with DrivAerNet++. The training set contains 5,819 designs, and the test set contains 1,154 designs, while the remaining designs are used for validation. Given the input vehicle geometry (e.g., point cloud), the model is trained to regress dense physical quantities at each surface point, which can subsequently be used to derive integrated aerodynamic quantities such as the drag coefficient.

\begin{table*}[!t]
  \caption{Detailed training configurations of GA-Field across datasets and tasks. All models are trained with the AdamW optimizer. “Warm up” denotes the number of iterations during which the learning rate is linearly increased from 0 to its initial value. “Grid size” specifies the grid-pooling resolutions used at different stages (e.g., [0.06, 0.96] indicates that the grid size doubles from 0.06 to 0.96 across stages).}
  \label{tab: impl training}
  \begin{center}
    \begin{small}
      \begin{sc}
        \begin{tabular}{l|l|cccccc}
            \toprule
            Dataset & Task & Input dim & LR & Warm up & Batch size & Epoch & Grid size \\
            \midrule
            \multirow{3}{*}{DrivAerNet++} & Pressure        & 7 & \multirow{3}{*}{$10^{-4}$} & \multirow{3}{*}{3000} & 4 & 200 & \multirow{3}{*}{[0.06,0.96]}\\
                                          & WallShearStress & 7 &  &  & 4 & 200 & \\                     
                                          & Velocity        & 6 &  &  & 4 & 200 & \\
            \midrule
            \multirow{3}{*}{DrivAerNet++} & OOD Setting 1   & 7 & \multirow{3}{*}{$10^{-4}$} & \multirow{3}{*}{800} & {2} & 100 & \multirow{3}{*}{[0.06,0.96]}\\
                                          & OOD Setting 2   & 7 &  &  & 4 & 100 & \\
                                          & OOD Setting 3   & 7 &  &  & 4 & 200 & \\
            \midrule
            \multirow{2}{*}{ShapeNet-Car} & Pressure        & 7 & \multirow{2}{*}{$10^{-4}$} & \multirow{2}{*}{800} & 2 & 400 & {[0.06,0.96]}\\
                                          & Velocity        & 7 &  &  & 2 & 400 & {[0.05,0.80]}\\
            \bottomrule
        \end{tabular}
      \end{sc}
    \end{small}
  \end{center}
  \vskip -0.1in
\end{table*}

\textbf{ShapeNet-Car.} ShapeNet-Car \cite{shapnetcar} is a CFD benchmark for learning-based aerodynamic modeling and drag coefficient estimation on synthetic vehicle shapes. It includes 889 simulations built from the ShapeNet “car” category \cite{chang2015shapenet}, where each design is simulated under a fixed uniform inlet wind speed of 20 m/s (72 km/h). For each case, the flow domain is discretized into an unstructured mesh with 32,186 points, and the dataset provides point-wise physical quantities in the surrounding flow, together with surface pressure and the resulting drag coefficient for supervision. Following protocols used by 3D-GeoCA \cite{ijcai2024p640} and Transolver \cite{wu2024Transolver}, we use 789 samples for training and 100 samples for testing.

\subsection{Training Settings}

The detailed training configurations for all datasets and tasks are summarized in Table~\ref{tab: impl training}. We adopt a unified architecture and train task-wise models independently. For the surface-field tasks on DrivAerNet++ and ShapeNet-Car, we use a 7D input feature consisting of 3D coordinates (3), surface normals (3), and the inflow-angle feature (1) defined in Sec.~\ref{main exp}. For flow-domain tasks, we use a 6D input on DrivAerNet++, formed by 3D coordinates (3) and the initial-velocity feature (3). For ShapeNet-Car flow prediction, we use a 7D input by further appending the signed distance function (SDF) (1) to the 6D representation, i.e., coordinates (3) + initial velocity (3) + SDF (1), where SDF encodes the geometry-to-flow distance field.

\subsection{Evaluation Metrics}
We evaluate the prediction $\hat{y}_i$ against the ground truth $y_i$ over $N$ test points. In our experiments, all metrics are computed on the test set and reported as the average over samples. 

\textbf{Mean Squared Error (MSE).} MSE measures the average squared deviation and assigns a larger penalty to large errors:
\begin{equation}
\mathrm{MSE}=\frac{1}{N}\sum_{i=1}^{N}\left(y_i-\hat{y}_i\right)^2.
\end{equation}

\textbf{Mean Absolute Error (MAE).} MAE computes the average absolute deviation and is less influenced by occasional large errors:
\begin{equation}
\mathrm{MAE}=\frac{1}{N}\sum_{i=1}^{N}\left|y_i-\hat{y}_i\right|.
\end{equation}

\textbf{Maximum Absolute Error (MaxAE).} MaxAE reports the worst-case absolute error among all samples:
\begin{equation}
\mathrm{MaxAE}=\max_{1\le i\le N}\left|y_i-\hat{y}_i\right|.
\end{equation}

\textbf{Coefficient of Determination ($R^2$).} $R^2$ quantifies how much of the variance in the ground truth is explained by the predictions:
\begin{equation}
R^2 = 1-\frac{\sum_{i=1}^{N}\left(y_i-\hat{y}_i\right)^2}{\sum_{i=1}^{N}\left(y_i-\bar{y}\right)^2}, \quad
\bar{y}=\frac{1}{N}\sum_{i=1}^{N}y_i.
\end{equation}

\textbf{Relative $\mathbf{L}_2$ Error.} Relative $\mathrm{L}_2$ error normalizes the Euclidean prediction error by the magnitude of the ground truth:
\begin{equation}
\mathrm{Rel}\,\mathrm{L}_2=\frac{\left\|\hat{\mathbf{y}}-\mathbf{y}\right\|_2}{\left\|\mathbf{y}\right\|_2}.
\end{equation}

\textbf{Relative $\mathbf{L}_1$ Error.} Relative $\mathrm{L}_1$ error normalizes the absolute error by the $\mathrm{L}_1$ magnitude of the ground truth:
\begin{equation}
\mathrm{Rel}\,\mathrm{L}_1=\frac{\left\|\hat{\mathbf{y}}-\mathbf{y}\right\|_1}{\left\|\mathbf{y}\right\|_1}.
\end{equation}


\end{document}